\documentclass[conference]{IEEEtran}

\usepackage{xurl}
\usepackage{amsmath}
\usepackage{amsfonts}
\usepackage{booktabs}
\usepackage{makecell}
\usepackage{array}
\usepackage{multirow}
\usepackage{graphicx}
\usepackage{mycommands}
\usepackage{forest}
\usepackage{comment}
\usepackage{tabularx}
\usepackage{orcidlink}
\usepackage{xspace}
\usepackage{float}
\hypersetup{hidelinks}

\usepackage{multirow}

\usepackage[table]{xcolor}

\newcounter{fr}
\renewcommand{\thefr}{FR-\arabic{fr}}

\newcommand{\fr}[1]{%
  \refstepcounter{fr}%
  \noindent\textbf{\thefr: #1}\quad
}

\newcounter{nr}
\renewcommand{\thenr}{NR-\arabic{nr}}

\newcommand{\nr}[1]{%
  \refstepcounter{nr}%
  \noindent\textbf{\thenr: #1}\quad
}

\usepackage{tikz}
\usetikzlibrary{arrows.meta,positioning,shapes,trees}
\tikzset{
  decision/.style = {rectangle, rounded corners, draw, thick, align=center, inner sep=4pt},
  chance/.style   = {circle, draw, thick, align=center, inner sep=1.5pt},
  terminal/.style = {diamond, draw, thick, aspect=2, inner sep=2pt, align=center},
  edge/.style     = {->, >=Stealth, thick},
  labelnode/.style= {font=\small, inner sep=1.5pt}
}

\begin{document}

\title{\fwname: Security Analysis and Mission-Aware Firewall for the CubeSat Space Protocol}

\author{
\IEEEauthorblockN{Owen Ramsey}
\IEEEauthorblockA{New Mexico State University\\
owenr@nmsu.edu}
\and
\IEEEauthorblockN{Carlos Gonzalez-Cortes}
\IEEEauthorblockA{Universidad de Santiago de Chile \\
carlos.gonzalez.c@usach.cl}
\and
\IEEEauthorblockN{Efrén López-Morales}
\IEEEauthorblockA{New Mexico State University\\
elopezm@nmsu.edu}
}

\maketitle

\begin{abstract}
The CubeSat Space Protocol (CSP) is widely used for communication in CubeSat missions, yet its security model and attack surface remain largely unexplored. In this work, we present a systematic security analysis of CSP and identify fundamental limitations in its trust assumptions, coarse-grained network abstractions, privileged management services, application command semantics, and lack of mission-context awareness. Based on these findings, we develop a taxonomy of CSP attacks spanning four classes: network identity abuse, management service abuse, application semantic abuse, and mission context abuse. We then introduce \fwname, a modular, mission-aware firewall that enforces CSP-specific policies derived from this taxonomy, including node and service authorization, command-level filtering, and mission-context constraints. We implement \fwname\  and evaluate it on a virtual testbed based on a real-world CubeSat mission and flight software. \fwname\ successfully blocks all \exploits representative attacks across the four attack classes while introducing only 1.42~ms of additional median round-trip latency. Our results demonstrate that protocol- and mission-aware policy enforcement provides a practical approach for securing CSP-based spacecraft networks.

\end{abstract}


\section{Introduction}

Modern CubeSats increasingly rely on reusable networking stacks and modular flight software for spacecraft development. Among these, the CubeSat Space Protocol (CSP) has emerged as one of the most adopted communication protocols for CubeSat missions, providing lightweight routing, and management services for communication between onboard subsystems and ground stations. CSP has been incorporated into open-source flight software~\cite{gonzalez2019architecture,suchai_flight_software_v2}, commercial spacecraft platforms~\cite{christiansen2011cubesat,planta2026satbleed}, and operational mission control systems~\cite{eoportal_gomx5,dlr2024_csp_egscc}, making it an important component of the small satellite ecosystem.

Historically, satellite systems relied on security by obscurity, e.g., limited public documentation and proprietary implementations~\cite{willbold2023space}. Today, however, the advent of commercial off-the-shelf (COTS) spacecraft hardware, open-source flight software, and accessible ground station infrastructure has substantially lowered the barrier to understanding satellite communication systems~\cite{lopezmorales2026honeysat}. Consequently, the security of widely deployed communication protocols such as CSP has become increasingly important.

Prior work has shown that CSP implementations contain vulnerabilities~\cite{NVD_CVE_2016_8596,NVD_CVE_2016_8597} and has introduced CPS-focused defenses such as Intrusion Detection Systems (IDS)~\cite{driouch2024distributed}.

However, despite CSP's widespread adoption, its security has not been systematically studied. Specifically, the protocol's own security model, attack surface, and inherent security limitations remain largely unexplored.

To address this gap, \emph{we perform the first systematic security analysis of CSP}. Specifically, we characterize the protocol's threat model and attack surface, providing a systematic assessment of the protocol from a security perspective.

Our analysis demonstrates that CSP's security challenges extend beyond software vulnerabilities and instead arise from fundamental limitations in the protocol's design. In particular, CSP provides only coarse-grained communication abstractions based on node identifiers and service ports, while the semantics of spacecraft telecommands are encoded within application-layer payloads. Furthermore, the protocol assumes trusted participants and provides no native mechanism for enforcing mission-specific security policies. Consequently, many security-critical operations cannot be authorized using CSP's native abstractions alone.

For example, the Space Attack Research and Tactic Analysis (SPARTA) framework, which provides a knowledge base of adversarial techniques and countermeasures for spacecraft systems, includes the \emph{Secure Command Modes} countermeasure (CM0055), which recommends constraining spacecraft commands based on conditions such as command source and geographic location~\cite{sparta_cm0055}. However, CSP does not provide mechanisms for enforcing such command policies.

Based on our analysis' findings, we developed \emph{a taxonomy of CSP attacks} that organizes protocol threats into four classes: network identity abuse, management service abuse, application semantic abuse, and mission context abuse. Rather than classifying individual exploits, the taxonomy captures the fundamental protocol abstractions that attackers can abuse and serves as the foundation for the design of our defense.

Motivated by our analysis' results, we propose \emph{\fwname}, a mission-aware firewall for CSP networks. Unlike traditional network firewalls, which enforce policies using IP addresses, and transport-layer ports, \fwname\ is designed to understand and secure CSP-specific communications. 

We designed \fwname\ to meet both functional and non-functional requirements. The functional requirements are based on the proposed attack taxonomy and define the security policies needed to mitigate each attack class. The non-functional requirements address deployment concerns, including modularity and mission configurability, enabling the firewall to be adapted across diverse CSP-based missions.

To evaluate \fwname, we developed a virtual testbed based on a real-world CubeSat mission and implemented \exploits representative CSP attack scenarios using a real-world flight software.

Our results show that \fwname successfully mitigates all \exploits attack scenarios across all four attack categories in our taxonomy while introducing only 1.42 ms of additional median packet latency, demonstrating that mission-aware CSP security can be achieved with limited runtime overhead.

In summary, this paper makes the following \textbf{contributions}:
\begin{itemize}

\item \textbf{Security Analysis:} We present, to the best of our knowledge, the first systematic security analysis of the CubeSat Space Protocol (CSP), characterizing its threat model, attack surface, and fundamental security limitations.

\item \textbf{Attack Taxonomy:} We develop the first taxonomy of CSP attacks, organizing threats according to the protocol abstractions they abuse rather than individual exploits.

\item \textbf{CSP Firewall:} We design and implement \fwname, a modular, mission-aware firewall whose functional requirements are based on the proposed taxonomy, enabling policy enforcement based on CSP identities, management services, application semantics, and mission context.

\item \textbf{Virtual Testbed Evaluation:} We evaluate \fwname\ on a virtual testbed based on a real-world CubeSat mission and representative attacks implemented on real flight software and measure the latency overhead \fwname introduces.

\end{itemize}

In the spirit of open and reproducible science, this paper's source code and experimental results are available online\footnote{https://github.com/cps-sec/stargate-csp-firewall}.

\section{Background}

In this section we lay out key concepts including the CubeSat Space Protocol (CSP) (Sec.~\ref{background:csp}), the CSP Management Protocol (CMP) (Sec.~\ref{subsec:cmp}), CSP adoption (Sec.~\ref{background:csp-fs}), and a Representative CSP Mission Network (Sec.~\ref{background:csp-network}).

\begin{figure*}
    \centering
    \includegraphics[width=1\linewidth]{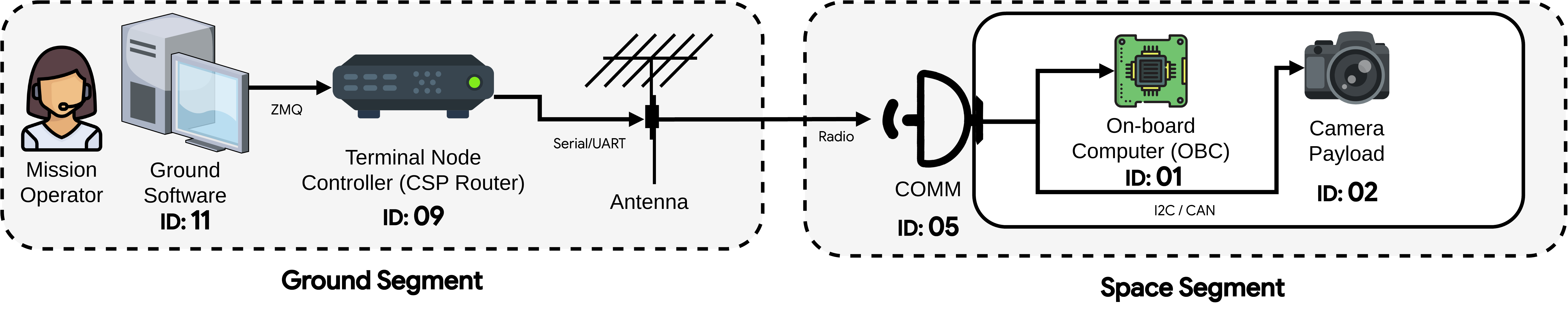}
    \caption{CSP network architecture of the real-world SUCHAI CubeSat mission~\cite{suchai_flight_software_v2}.}
    \label{fig:suchai-mission}
\end{figure*}

\subsection{The CubeSat Space Protocol}\label{background:csp}

The CubeSat Space Protocol (CSP) is a lightweight protocol designed for resource-constrained satellite systems~\cite{libcsp_main}. CSP provides routing, transport, management services, and interface drivers, forming a complete communication stack for both ground and space segment communications.

CSP follows a packet-based design with a compact header format that encodes source and destination addresses, ports, and control flags. The main CSP implementation, \emph{libcsp}~\cite{libcsp_main}, uses a header optimized for low overhead and efficient processing on embedded hardware. This design makes CSP well-suited for microcontrollers and real-time operating systems commonly used in space systems.

Two major versions of CSP are supported by \emph{libcsp}~\cite{libcsp_main}. We focus our implementation and evaluation on CSP 1, which is used by our target flight software and testbed. While CSP 2 introduces changes to the protocol header, the security limitations examined in this work largely arise from CSP's communication and authorization model and are therefore not specific to the CSP 1 header format.

\textbf{CSP Communication Model.} CSP follows a service-oriented communication model in which each spacecraft subsystem is assigned a unique \emph{node identifier (ID)} and exposes its functionality through one or more \emph{ports}~\cite{libcsp_network_topology}. Applications, such as telemetry, and network management, communicate by exchanging CSP packets addressed to a destination node and port. Upon receiving a packet, a \emph{CSP router} first determines whether the packet should be forwarded or delivered locally based on the destination node ID. If the packet is destined for the local node, it is dispatched to the associated port, otherwise it is forwarded to a default route or interface.

\textbf{CSP Routing.} CSP employs a distributed routing architecture in which every CSP-enabled node can function as both a communication endpoint and a router. Each node maintains its own routing table and forwards packets toward their destination. This decentralized design improves flexibility but provides no centralized enforcement point where communication policies can be applied~\cite{libcsp_network_topology}.

\textbf{CSP Interfaces.} CSP is transport-independent and can operate over multiple communication interfaces. Common interfaces include Controller Area Network (CAN), Inter-Integrated Circuit (I\textsuperscript{2}C), serial links, and ZeroMQ (ZMQ)~\cite{libcsp_protocol_stack}.

%Routing can be configured statically or dynamically, enabling multi-hop communication across distributed spacecraft architectures and ground infrastructure.

\textbf{CSP Security Features.} CSP provides a limited set of security mechanisms in its commonly used implementations, mainly XTEA and HMAC-SHA1 for encryption and authentication. However, authentication, integrity, and confidentiality are not enforced at the protocol level, and are instead left to higher-layer mechanisms or mission-specific extensions. Furthermore, XTEA and HMAC-SHA1 are fundamentally weak by modern cryptographic standards~\cite{libcsp_main}.

Recent efforts toward CSP version 2 aim to introduce optional security features. However, these enhancements are not yet widely deployed in operational systems.

\begin{comment}
    
\carlos{talk about CSP security features}

\carlos{amateur frequency requires unencrypted downlink traffic}

In this work, we consider a CSP-based communication architecture representative of real-world deployments, using LibCSP as the underlying implementation. This allows us to study security challenges in practical settings and motivates the need for external enforcement mechanisms, such as the CSP-aware firewall proposed in this paper.

\end{comment}

\subsection{CSP Management Protocol (CMP)}
\label{subsec:cmp}

CSP uses service ports to identify functionality exposed by a node. Of the 64 possible CSP ports, ports 0--7 are reserved for standardized CSP services, while the remaining ports are available for mission-specific applications~\cite{libcsp_network_topology}. These standardized services provide functionality for network management, diagnostics, and spacecraft operation, including node identification, interface statistics, memory access, clock synchronization, routing configuration, and system reboot~\cite{gomspace_nanoground_manual}.

Although these services support legitimate spacecraft operations, several expose security-sensitive functionality when accessed by an unauthorized node. For example, \texttt{PEEK} and \texttt{POKE} provide memory access, \texttt{ROUTE\_SET} modifies routing information, \texttt{CLOCK} modifies system time, and the reboot service can restart a CSP node. %Table~\ref{tab:cmp_services} in Appendix~\ref{app:cmp} summarizes the standardized CSP services, their intended purpose, and their potential security implications.

\subsection{CSP Adoption in Operational Missions}\label{background:csp-fs}

The CubeSat Space Protocol is widely adopted and has been integrated into multiple open-source and commercial flight software stacks, including the SUCHAI Flight Software~\cite{suchai_flight_software_v2}, and is supported by organizations such as the German Space Operations Center~\cite{dlr2024_csp_egscc}. Importantly, GomSpace, a spacecraft manufacturer has adopted CSP as their de facto standard for most of their hardware~\cite{gomspaceA3200,christiansen2011cubesat}.

To better understand the extent of CSP adoption, we analyzed publicly available satellite information from SatNOGS~\cite{satnogs} and the nanosats database~\cite{nanosats_database}. Our analysis identified 104 operational satellites that use CSP for communications in some capacity. For example, several missions use GomSpace's NanoCom AX.100 transceiver which implements CSP communications by default~\cite{gomspace_nanoground_manual}. Our CSP-enabled satellite missions dataset is available online\footnote{https://github.com/cps-sec/csp-satellites-list}.

\subsection{Representative CSP Mission Network}
\label{background:csp-network}

To ground our research in a realistic deployment, we consider the network architecture of the SUCHAI operational CubeSat mission that employs CSP~\cite{suchai_flight_software_v2}. 

Fig.~\ref{fig:suchai-mission} depicts the CSP mission network. The architecture is divided into a \emph{ground segment} and a \emph{space segment}, which communicate through a radio link. The ground segment consists of the mission operator, the ground software, a Terminal Node Controller (TNC), and the ground antenna. The ground software generates CSP packets that are forwarded to the TNC, which interfaces with the radio hardware and transmits the packets over the wireless link. The TNC also acts as a CSP router using a predefined CSP routing table.

The space segment consists of the spacecraft's On-Board Computer (OBC) and one or more onboard subsystems, such as a camera payload. The OBC executes the CSP stack and serves as the primary communication gateway for the spacecraft. Upon receiving a CSP packet from the radio subsystem, the OBC routes the packet to the destination subsystem based on the destination node identifier and service port. Each subsystem is assigned a unique CSP node identifier and exposes its functionality through one or more CSP services.

\section{CubeSat Space Protocol Security Analysis}
\label{sec:analysis}

To systematically identify the security challenges introduced by the CubeSat Space Protocol (CSP), we first analyze its communication model, management services, and operational assumptions from a security perspective. Specifically, we (i) define the CSP threat model (Sec.~\ref{subsec:threat-model}), (ii) the protocol attack surface (Sec.~\ref{subsec:attack-surface}), and (iii) develop a taxonomy of representative CSP attacks based on the identified attack vectors and trust boundaries (Sec.~\ref{subsec:taxonomy}).

\subsection{Threat Model}\label{subsec:threat-model}

\textbf{Attacker capabilities.}
We consider a CSP-based satellite mission in which Stargate is
deployed as a policy-enforcement point on the authorized communication
path between the ground segment and spacecraft. We assume that CSP
traffic reaching the spacecraft through this path traverses Stargate.
The adversary can generate arbitrary CSP packets through the authorized
ground communication infrastructure, for example through compromised
ground software, operator misuse, an insider, or a secondary ground
system connected to the mission network. The attacker possesses
knowledge of valid CSP node identifiers, service ports, and publicly
documented management services, enabling the construction of
syntactically correct CSP packets targeting spacecraft subsystems.

\textbf{Attacker objectives.} The attacker's objective is to invoke unauthorized spacecraft functionality, disrupt mission operations, or access sensitive information by abusing legitimate CSP services. Representative attacks include unauthorized invocation of CSP Management Protocol (CMP) services, misuse of application-specific telecommands, and denial-of-service through excessive command transmission.

\textbf{Out of scope.}
We assume that CSP traffic using the protected ground-space
communication path traverses Stargate. Adversaries capable of bypassing
this enforcement point, for example by directly transmitting to the
spacecraft over an independent RF link, compromising communication
infrastructure downstream of Stargate, or compromising an onboard CSP
node, are outside the scope of this work. Attacks targeting the physical
layer, radio-frequency jamming, cryptographic mechanisms, key
management, or vulnerabilities in specific CSP implementations are
also outside the scope of this work.

\subsection{CSP Attack Surface}\label{subsec:attack-surface}

We focus on the CSP-specific attack surface, in particular, we go beyond TCP/IP network connectivity and includes network identifiers, ports, management services, application-layer commands, and mission context.

\textbf{Node Identifiers.}
Every CSP-enabled subsystem is assigned a unique node ID. An attacker capable of transmitting arbitrary CSP packets may target any reachable node, including critical spacecraft components such as the Command and Data Handling (C\&DH), or mission payloads.

\textbf{Service Ports.}
Each node exposes functionality through one or more CSP ports. While ports provide logical separation between services, they represent only coarse-grained access control. Once a packet is delivered to a service, CSP does not distinguish between authorized and unauthorized operations.

\textbf{Management Services.}
As discussed in Sec.~\ref{subsec:cmp}, CSP includes the CSP Management Protocol (CMP), which provides standardized services such as routing. These services are intended for spacecraft operation but also expose functionality that could be abused. For example, CMP routing features can be abused for route poisoning.

\textbf{Application Command Semantics.}
Mission functionality is typically implemented through application-specific telecommands whose semantics are encoded within the packet payload rather than the CSP header. For example, two packets may have the same source ID, destination ID, and service port, yet one requests a temperature reading while the other initiates a firmware update. Since CSP only exposes the node identifiers and service ports, it cannot distinguish between these fundamentally different operations, preventing authorization decisions based solely on CSP headers.

\textbf{Mission Context.}
The legitimacy of many spacecraft commands depends on the current operational context of the mission. Commands that are valid during one mission phase may be unsafe or unauthorized during another. Examples include payload activation outside imaging windows, or maintenance operations. These constraints are external to the CSP protocol and cannot be enforced by native CSP mechanisms.

\subsection{CSP Attack Taxonomy}\label{subsec:taxonomy}

\begin{figure}
    \centering
    \includegraphics[width=1\linewidth]{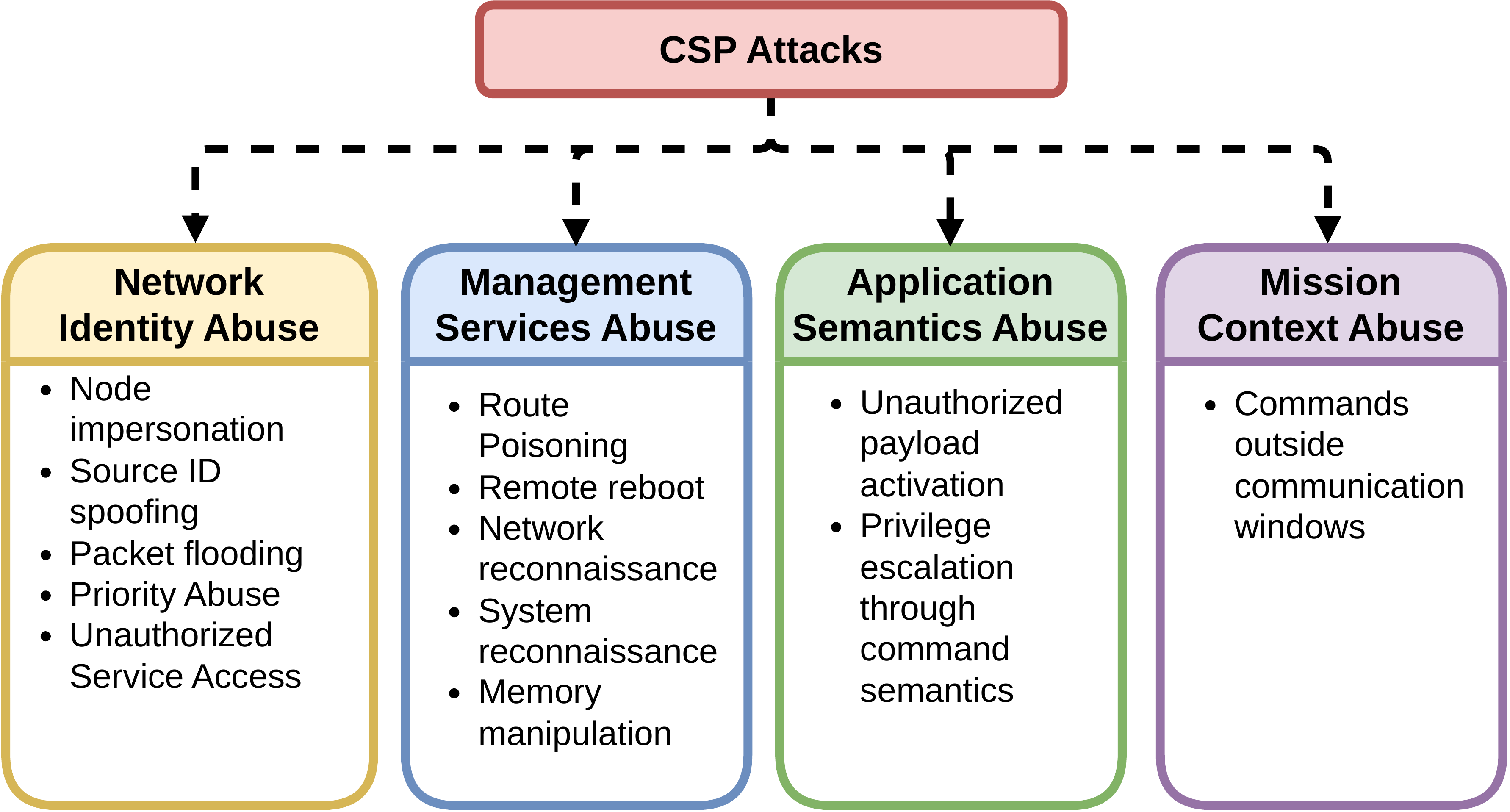}
    \caption{CSP Attack Taxonomy}
    \label{fig:taxonomy}
\end{figure}

Based on the attack surface described in the previous section, we derive a taxonomy of attacks against CSP-based spacecraft networks depicted in Fig.~\ref{fig:taxonomy}. Rather than classifying attacks according to specific exploits, we categorize them according to the protocol abstraction they abuse.

\textbf{Network Identity Abuse.}
The first class of attacks exploits CSP network identifiers, including source and destination node IDs, service ports, and packet priorities. Since CSP assumes that participating nodes are trusted, an attacker capable of injecting CSP packets may perform \emph{node impersonation} by spoofing the source ID of a legitimate node, \emph{unauthorized node communication} by sending packets to spacecraft subsystems that should not be reachable, \emph{service enumeration} by probing ports to discover exposed applications, \emph{unauthorized service access} by invoking privileged CSP or application services, \emph{priority abuse} by marking malicious traffic as high priority to compete with legitimate communications, or \emph{traffic flooding} by overwhelming critical nodes with excessive packets. These attacks exploit the lack of protocol-level authentication and
authorization rather than implementation vulnerabilities. In particular,
because CSP node identifiers are not cryptographically authenticated,
a source ID alone cannot establish the identity of the sender.

\textbf{Management Service Abuse.}
The second class targets the CSP Management Protocol (CMP). Representative attacks include \emph{information disclosure} by reading sensitive memory with \texttt{PEEK}, \emph{memory manipulation} by modifying runtime state with \texttt{POKE}, \emph{route poisoning} by altering packet forwarding paths through \texttt{ROUTE\_SET}, \emph{time manipulation} by changing the spacecraft clock to disrupt mission scheduling, and \emph{remote reboot} by forcing an unexpected system restart. These attacks exploit privileged management functionality intended for trusted operators.

\textbf{Application Semantic Abuse.}
The third class targets application-level command semantics. Many spacecraft applications expose multiple telecommands through a single CSP service port, with the requested operation encoded within the application payload rather than the CSP header. Representative attacks include \emph{firmware update abuse} by invoking software update commands through an otherwise authorized service, \emph{unauthorized subsystem control} by issuing reset, shutdown, or mode transition commands, and \emph{payload activation} by enabling mission payloads without operator authorization. These attacks exploit CSP's inability to distinguish between different application commands based solely on node identifiers and service ports.

\textbf{Mission Context Abuse.}
The final class targets mission-specific operational context. Unlike semantic attacks, the command itself may be legitimate, but its execution violates mission constraints. Representative attacks include \emph{out-of-window command execution} by transmitting telecommands outside authorized ground contact periods, \emph{mission-phase violations} by issuing deployment or calibration commands during nominal operations, and \emph{state-dependent command abuse} by activating payloads or changing operating modes while the spacecraft is in a protected state such as safe mode. Since these policies depend on the spacecraft's operational state rather than the CSP protocol, they cannot be enforced using native CSP mechanisms alone.

\subsection{Representative Attack Scenarios}

The following attack scenarios illustrate how each category of the proposed taxonomy may be exploited in practice.

\textbf{Rogue CSP Node (Network Identity Abuse).}
A rogue CSP node joins the network and attempts to communicate with the spacecraft without authorization. By targeting valid spacecraft node IDs and service ports, the attacker can issue unauthorized requests, probe exposed services, or flood critical subsystems with malicious traffic. Since CSP assumes that participating nodes are trusted, native CSP mechanisms cannot distinguish legitimate operators from unauthorized network participants.

\textbf{ Route Poisoning (Management Service Abuse).}
An attacker abuses the CMP \texttt{ROUTE\_SET} service to modify the spacecraft's routing table. By redirecting traffic to a rogue node, creating invalid routes, or blackholing legitimate communication paths, the attacker can intercept, misroute, or disrupt CSP communications between spacecraft subsystems and the ground segment. Since \texttt{ROUTE\_SET} is intended for trusted network reconfiguration, unauthorized access to this service enables direct manipulation of the spacecraft's communication topology.

\textbf{Unauthorized Command on an Authorized Port (Application Semantic Abuse).}
An attacker sends a privileged telecommand through an otherwise authorized CSP service port. For example, the same application port may accept both a benign telemetry request and a firmware update command. Although the CSP headers are identical, the requested operations have vastly different security implications. Since the command is encoded within the application payload rather than the CSP header, CSP cannot distinguish between these operations using node identifiers and service ports alone.

\textbf{Command Outside an Authorized Communication Window (Mission Context Abuse).}
An attacker transmits an otherwise legitimate telecommand outside an authorized ground contact window. Although the command is valid, it violates mission-specific communication policies and should not be accepted. Since communication windows are external to the CSP protocol, native CSP mechanisms cannot enforce these temporal constraints.

\subsection{Security Analysis Summary}

Our security analysis demonstrates that securing CSP networks requires more than traditional packet filtering based on network addresses and ports. CSP security risks arise at multiple layers, including network identities, privileged management services, application command semantics, and mission-specific operational constraints. Consequently, effective protection requires a policy enforcement mechanism capable of authorizing \emph{who} may communicate, \emph{which} protocol services may be accessed, \emph{what} application commands may be executed, and \emph{when} those commands are permitted according to mission context. These observations directly motivate the design of \fwname, a CSP mission policy-aware firewall which we now describe.

\section{\fwname\ Firewall Design}
\label{sec:design}

Traditional network firewalls are designed for Internet Protocol (IP) networks, where security policies are enforced using packet attributes such as source and destination IP addresses, transport-layer ports, and protocol identifiers. These abstractions are effective for conventional enterprise networks because application services are exposed through well-defined TCP or UDP ports. In contrast, spacecraft systems using the CubeSat Space Protocol (CSP) employ a different communication model based on node identifiers, CSP service ports, management services, and mission-specific telecommands.

Furthermore, CSP traffic is commonly transported over communication technologies such as ZeroMQ tunnels, CAN buses, serial links, or radio frames. In these deployments, a conventional firewall only observes the underlying transport protocol and lacks visibility into CSP headers and application payloads. Consequently, it cannot enforce CSP-specific security policies, including restrictions on node identities, management services, application commands, or mission-dependent operational constraints.

To address these limitations, we propose \fwname, a policy-aware firewall specifically designed for CSP networks. Rather than relying on traditional IP semantics, \fwname\ inspects CSP packets and evaluates them against mission-defined policies derived from the security analysis presented in Sec.~\ref{sec:analysis}. The following subsections describe the design requirements and architecture of the proposed firewall.

\subsection{Design Requirements}

Based on the security analysis presented in the previous section, we follow the below requirements.

\subsubsection{Functional Requirements}

The following four functional requirements are derived directly from the CSP attack taxonomy (Sec.~\ref{subsec:taxonomy}) and define the core security capabilities that \fwname must provide.

\fr{Identity-aware Policy Enforcement}\label{sr:identity}
The firewall shall enforce communication policies based on CSP identity to restrict unauthorized network communication.

\fr{Service-aware Policy Enforcement}\label{sr:service}
The firewall shall selectively authorize privileged CSP management services to prevent management services abuse attacks.

\fr{Semantic-aware Policy Enforcement}\label{sr:semantic}
The firewall shall inspect application payloads to distinguish and authorize individual telecommands exposed through the same CSP service port to prevent application semantics abuse attacks.

\fr{Mission-aware Policy Enforcement}\label{sr:context}
The firewall shall enforce mission-specific operational policies, such as communication windows and mission phases, when authorizing telecommands to prevent mission context abuse attacks.

\subsubsection{Non-Functional Requirements}

The following non-functional requirements allow \fwname to be integrated into diverse CSP-based missions.

\nr{Modular Architecture}\label{nr:modular}
Unlike traditional network firewalls, CSP deployments vary significantly across missions in terms of communication interfaces, network topology, onboard applications, and supported telecommands. Therefore, \fwname\ has a modular architecture that allows mission-specific components, including CSP interfaces, protocol parsers, and policy modules, to be added, replaced, or extended without modifying the entire firewall.

\nr{Mission Configurability}\label{nr:mission} \fwname is designed to support mission-specific security policies, e.g., authorized node IDs, and other operational constraints.

\subsection{\fwname Firewall Architecture}

\begin{figure}[t]
    \centering
    \includegraphics[width=1.0\linewidth]{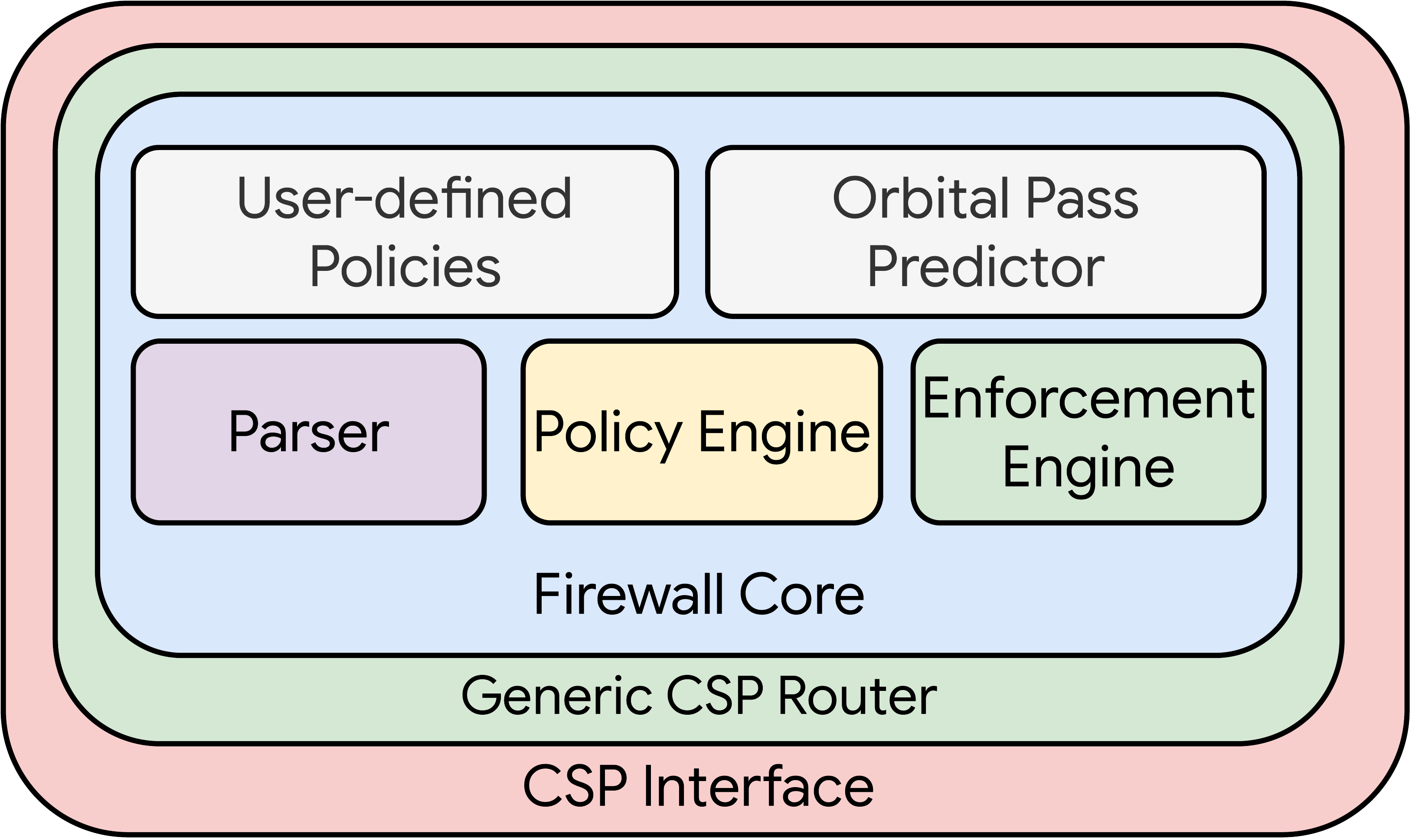}
    \caption{\fwname Firewall Modular  Architecture}
    \label{fig:architecture}
\end{figure}

Figure~\ref{fig:architecture} illustrates the architecture of \fwname. The firewall is deployed at the ground-space communication boundary, where it intercepts all CSP traffic before it reaches the spacecraft. Incoming packets first enter the \textit{CSP Interface}, which provides an abstraction over the underlying communication medium. This modular interface satisfies \ref{nr:modular} by allowing \fwname\ to be deployed across diverse CSP-based missions without modifying the core firewall logic.

Packets are then forwarded to the \textit{Generic CSP Router}, which receives and forwards CSP packets through the firewall processing pipeline. The \textit{Parser} extracts security-relevant information from each packet, including source and destination node identifiers, service ports, packet priorities, CMP operations, and application-specific telecommands. These attributes provide the input required for policy evaluation.

The extracted information is evaluated by the \textit{Policy Engine}, which implements the functional requirements derived from the CSP security analysis (Sec.~\ref{sec:analysis}). The policy engine evaluates packets against multiple classes of security policies, including network identity policies (\ref{sr:identity}), management service policies (\ref{sr:service}), application semantic policies (\ref{sr:semantic}), and mission context policies (\ref{sr:context}). Mission-specific policies are loaded from the \emph{User-defined Policies} file (\ref{nr:mission}), while mission timing constraints are supplied by the \textit{Orbital Pass Predictor}, enabling enforcement of communication windows.

Finally, the policy decision is passed to the \textit{Enforcement Engine}, which determines whether the packet should be forwarded, dropped, or logged.
\begin{figure}
    \centering
    \includegraphics[width=1\linewidth]{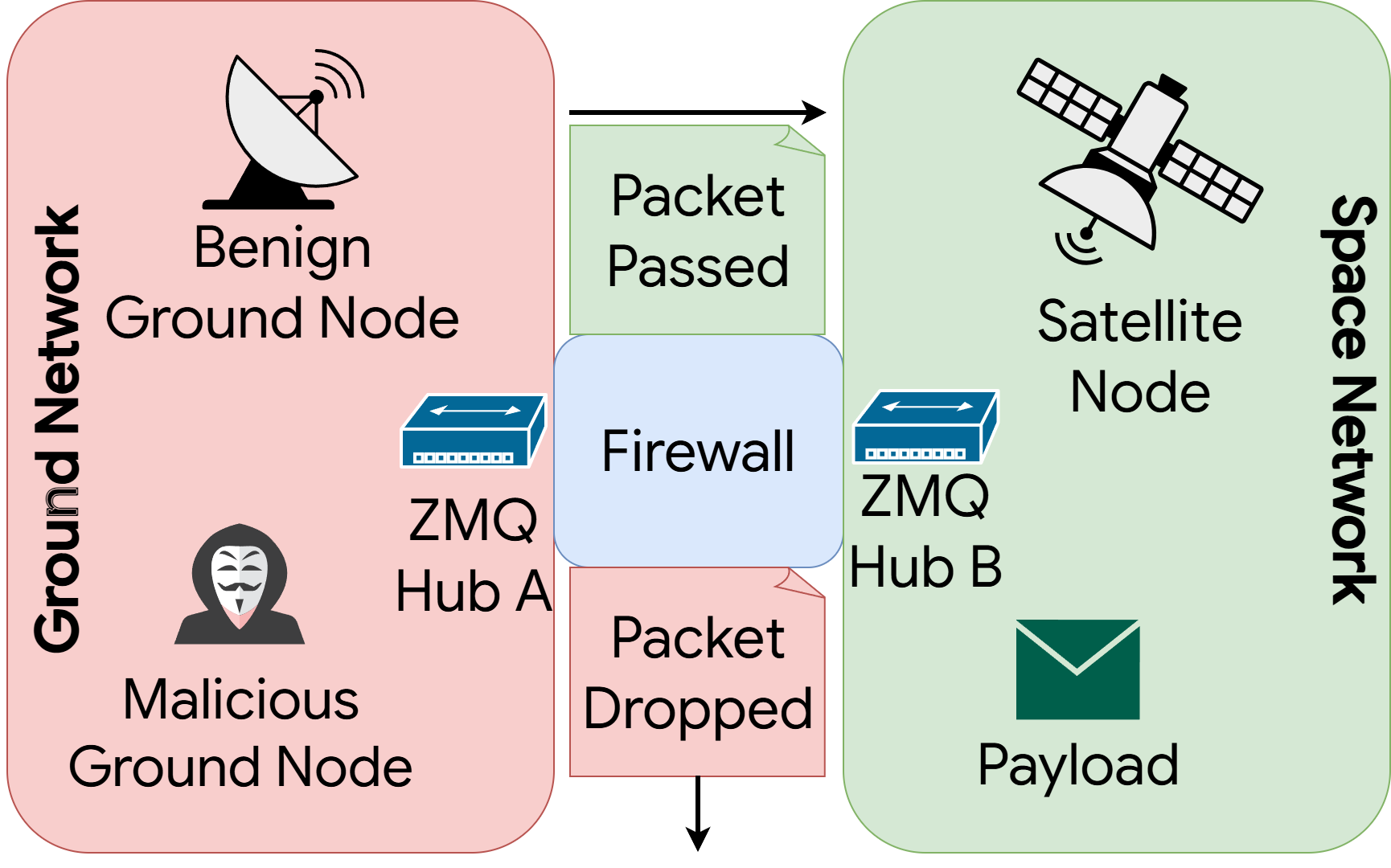}
    \caption{CSP Virtual Testbed}
    \label{fig:testbed}
\end{figure}

\subsection{Policy Engine Architecture}

\begin{figure}[H]
    \centering
    \includegraphics[width=1\linewidth]{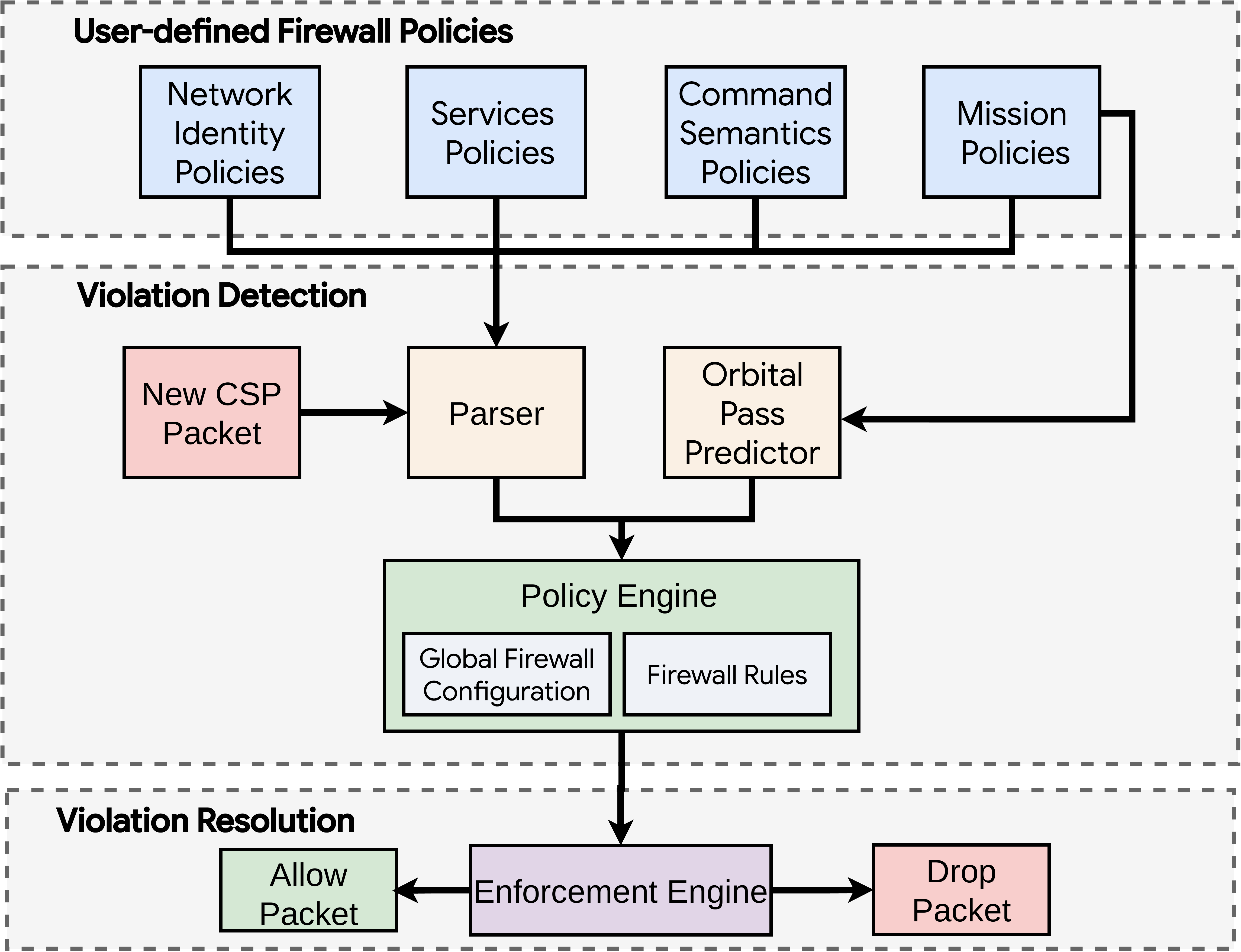}
    \caption{Policy Engine Diagram. Inspired by \cite{Hu2014Flowguard}}
    \label{fig:policy}
\end{figure}

The Policy Engine depicted in Fig.~\ref{fig:policy} is responsible for evaluating parsed CSP packets against mission-defined security policies. Unlike traditional firewalls that base decisions on network-layer attributes, \fwname\ performs multi-dimensional policy evaluation using the information extracted by the parser. Specifically, the Policy Engine evaluates each packet against four classes of policies derived from the CSP threat taxonomy and functional requirements. A packet is forwarded only if it satisfies all applicable policies; otherwise, it is rejected and logged.

\textbf{Network Identity Policies (\ref{sr:identity}).}
Network identity policies regulate communication based on CSP source and destination node identifiers, service ports, and packet priorities. These policies ensure that only authorized nodes are permitted to communicate with protected spacecraft services and prevent unauthorized or rogue CSP nodes from accessing mission-critical resources.

\textbf{Management Service Policies (\ref{sr:service}).}
Management service policies control access to privileged CSP Management Protocol (CMP) services. Operations such as \texttt{PEEK}, \texttt{POKE}, \texttt{ROUTE\_SET}, \texttt{CLOCK}, and \texttt{REBOOT} can be selectively authorized or denied according to mission requirements, preventing unauthorized administrative operations.

\textbf{Application Semantic Policies (\ref{sr:semantic}).}
Application semantic policies provide fine-grained authorization of spacecraft telecommands. Since multiple telecommands may share the same CSP service port, the parser extracts the application command from the packet payload before policy evaluation. This enables \fwname\ to distinguish between benign operations and privileged commands that are otherwise indistinguishable at the CSP protocol layer.

\textbf{Mission Context Policies (\ref{sr:context}).}
Mission context policies evaluate operational constraints that are external to the CSP protocol. Examples include communication windows, mission phases, and spacecraft operational modes. Consequently, a command that satisfies all network and application-level policies may still be denied if it violates mission-specific operational constraints.

\section{Implementation}

We implemented \fwname\ in C using the libcsp library~\cite{libcsp_main}. Security policies are specified in a YAML configuration file.

Mission context policies are implemented using the Skyfield library~\cite{skyfield_docs}, which computes satellite visibility from Two-Line Element (TLE) data. During runtime, the orbital pass predictor determines whether the spacecraft is within an authorized communication window and supplies this information to the policy engine. We use publicly available TLE data obtained from CelesTrak~\cite{celestrak}.

\subsection{CSP Virtual Testbed}
\label{subsec:testbed}
Based on the architecture of the real-world CubeSat mission discussed in Sec.~\ref{background:csp-network}, we developed the virtual CSP testbed shown in Fig.~\ref{fig:testbed}. The testbed includes five Docker containers that emulate the main components of a CSP-enabled mission. All nodes use \textbf{libcsp}. The \textbf{benign ground node} represents an authorized ground station that generates legitimate CSP commands, while the \textbf{malicious ground node} represents an adversary capable of transmitting crafted CSP packets toward the spacecraft. Both ground nodes connect to \textbf{ZMQ Hub A}, which represents the ground-side CSP network.

The \textbf{firewall} is positioned between the ground and space networks so that all CSP traffic destined for the spacecraft must traverse it. Packets permitted by the configured security policies are forwarded to \textbf{ZMQ Hub B}, while packets that violate a policy are dropped before entering the space network. ZMQ Hub B connects the firewall to the spacecraft-side CSP nodes. The \textbf{satellite node} runs \emph{real open-source spacecraft flight software} with CSP support, providing a realistic target for evaluating the firewall rather than a synthetic CSP application. The \textbf{payload} represents an additional onboard CSP subsystem communicating with the satellite node.

\section{Evaluation}

We evaluate \fwname\ by answering the following research questions:

\begin{itemize}
    \item \textbf{RQ1:} Can \fwname\ mitigate representative attacks
    from each class of the CSP attack taxonomy? (Sec.~\ref{subsec:attack-mitigation})
    
    \item \textbf{RQ2:} What runtime overhead does \fwname\ introduce
    to CSP communication? (Sec.~\ref{subsec:overhead})
\end{itemize}

\subsection{Experimental Setup}

We evaluate \fwname\ using the virtual CSP testbed depicted in Fig.~\ref{fig:testbed} and discussed in Sec.~\ref{subsec:testbed}.

Experiments were conducted in a VirtualBox virtual machine running
Ubuntu 25.04 with 8 virtual CPUs and 16~GB of RAM. For each attack
scenario, we first executed the attack without policy enforcement to
confirm its effect on the target CSP node. We then repeated the attack
with the corresponding \fwname\ policy enabled and recorded whether
the malicious packet was forwarded or dropped.

For mission-context experiments, we used Skyfield and publicly
available TLE data from CelesTrak to determine communication windows
between the spacecraft and a ground station. The resulting communication state was provided to \fwname's
policy engine to determine whether telecommands were authorized.

\subsection{Attack Mitigation Effectiveness}
\label{subsec:attack-mitigation}

To answer \textbf{RQ1} we evaluated whether \fwname\ can mitigate representative attacks
from each of the four classes identified by our CSP security analysis.
Table~\ref{tab:attack_evaluation} summarizes the evaluated attacks,
their effects, the policy mechanism used for mitigation, and the
observed result.

\subsubsection{Network Identity Abuse}

We evaluated four attacks targeting CSP network-level attributes:
rogue-node communication, unauthorized service access, priority abuse,
and packet flooding. For the rogue-node experiment, the legitimate
ground station (node 10) was permitted to communicate with the
spacecraft, while traffic originating from the malicious ground
station (node 11) was denied. \fwname\ correctly forwarded traffic
from the authorized node and dropped traffic originating from the
unauthorized node.

We similarly verified service-port and priority enforcement by
transmitting packets containing both authorized and restricted values.
Packets targeting restricted ports or using prohibited priorities were
dropped, while permitted packets were forwarded. Finally, the
rate-limiting policy restricted command transmission to five commands
per minute, preventing the malicious node from exceeding the
mission-defined  rate.

\subsubsection{Management Service Abuse}

We evaluated five security-sensitive CSP management operations:
\texttt{PEEK}, \texttt{POKE}, \texttt{ROUTE\_SET}, \texttt{CLOCK},
and \texttt{REBOOT}. Without policy enforcement, the malicious ground
node was able to invoke these operations against the target CSP node,
enabling memory disclosure, memory modification, routing-table
modification, system-time manipulation, and remote reboot.

With \fwname\ enabled, all five operations were blocked using
service- and payload-level policies. CMP operations sharing the same
service port were distinguished using their command semantics,
demonstrating that port-level filtering alone is insufficient for
protecting CSP management.

\subsubsection{Application Semantic Abuse}

To evaluate semantic-aware enforcement, we transmitted both an
authorized and a restricted telecommand through the same CSP service
port. Because the packets share the same CSP source, destination, and
service port, a policy operating exclusively on CSP header fields
cannot distinguish between the two operations.

\fwname\ parsed the application payload and identified the requested
telecommand. The authorized command was forwarded, while the
restricted command was dropped. This experiment demonstrates that
application-semantic inspection enables command-level authorization
beyond CSP's native node and port abstractions.

\subsubsection{Mission Context Abuse}

We evaluated mission-aware enforcement by transmitting the same
otherwise authorized telecommand both inside and outside an authorized
communication window. Communication windows were determined using the
orbital-pass predictor described in Sec.~\ref{sec:design}.

When the spacecraft was within an authorized communication window,
\fwname\ forwarded the command after it satisfied the remaining
policies. When the same command was transmitted outside the
communication window, \fwname\ dropped the packet. Thus, identical CSP
traffic can receive different authorization decisions based on
mission context that is not represented within the CSP packet itself.

Our evaluation focuses on communication-window policies as one
representative form of mission context; mission-phase and
spacecraft-state enforcement are not evaluated in the current prototype.

\begin{table*}[t]
\centering
\caption{Effectiveness of \fwname\ against representative CSP attacks
across the four attack classes in our taxonomy.}
\label{tab:attack_evaluation}

\renewcommand{\arraystretch}{1.18}
\setlength{\tabcolsep}{5pt}

\begin{tabularx}{\textwidth}{
    >{\raggedright\arraybackslash}m{2.5cm}
    >{\raggedright\arraybackslash}m{3.0cm}
    >{\raggedright\arraybackslash}X
    >{\raggedright\arraybackslash}m{3.0cm}
    >{\centering\arraybackslash}m{1.25cm}}
\toprule
\textbf{Attack Class} &
\textbf{Attack} &
\textbf{Effect Without \fwname} &
\textbf{Enforcement Policy} &
\textbf{Result} \\
\midrule

% =========================================================
% Network Identity Abuse
% =========================================================
\multirow{4}{2.5cm}{\raggedright\textbf{Network Identity Abuse}}
& Rogue node
& Unauthorized node sends spacecraft commands
& Node identity
& Blocked \\

\cmidrule(l){2-5}

& Unauthorized service access
& Accesses a restricted CSP service
& Service port
& Blocked \\

\cmidrule(l){2-5}

& Priority abuse
& Sends traffic using a restricted priority
& Packet priority
& Blocked \\

\cmidrule(l){2-5}

& Packet flooding
& Exceeds the permitted command rate
& Rate limiting
& Blocked \\

\midrule

% =========================================================
% Management Service Abuse
% =========================================================
\multirow{5}{2.5cm}{\raggedright\textbf{Management Service Abuse}}
& Memory disclosure (\texttt{PEEK})
& Reads memory from the target node
& CMP semantics
& Blocked \\

\cmidrule(l){2-5}

& Memory modification (\texttt{POKE})
& Writes attacker-controlled data to memory
& CMP semantics
& Blocked \\

\cmidrule(l){2-5}

& Route poisoning (\texttt{ROUTE\_SET})
& Modifies CSP routing information
& CMP semantics
& Blocked \\

\cmidrule(l){2-5}

& Time manipulation (\texttt{CLOCK})
& Modifies the spacecraft system clock
& CMP semantics
& Blocked \\

\cmidrule(l){2-5}

& Remote reboot (\texttt{REBOOT})
& Forces the target CSP node to restart
& Service port
& Blocked \\

\midrule

% =========================================================
% Application Semantic Abuse
% =========================================================
\textbf{Application Semantic Abuse}
& Unauthorized telecommand
& Executes a restricted command on an authorized port
& Command semantics
& Blocked \\

\midrule

% =========================================================
% Mission Context Abuse
% =========================================================
\textbf{Mission Context Abuse}
& Out-of-window command
& Executes a valid command outside an authorized contact window
& Mission context
& Blocked \\

\bottomrule
\end{tabularx}
\end{table*}

\subsection{Overhead}
\label{subsec:overhead}

\begin{table}
\centering
\caption{\fwname\ Round-Trip Time (RTT) Overhead}
\label{tab:firewall_results}
\begin{tabular}{lcccc}
\toprule
Metric & No Firewall (ms) & Firewall (ms) & $\Delta$ (ms) & Change \\
\midrule
Min    & 3.00   & 3.50   & +0.50   & +16.67\% \\
Max    & 603.33 & 900.00 & +296.67 & +49.17\% \\
Mean   & 51.76  & 63.36  & +11.60  & +22.41\% \\
Median & 10.83  & 12.25  & +1.42   & +13.11\% \\
Mode   & 6.67   & 8.67   & +2.00   & +29.99\% \\
\bottomrule
\end{tabular}
\end{table}

To answer \textbf{RQ2} we evaluate the overhead introduced by \fwname, we measured the round-trip time (RTT) of CSP \texttt{ping} messages with and without the firewall enabled. A total of 100 RTT measurements were collected per experiment and repeated five times. Table~\ref{tab:firewall_results} reports the average minimum, maximum, mean, median, and mode across all experiments.

The results indicate that \fwname introduces a median RTT increase
of 1.42 ms and a mean increase of 11.60 ms. The difference between
the median and mean, together with the observed maximum RTTs,
indicates substantial variability in the virtualized measurement
environment. Therefore, these measurements should be interpreted as
an initial characterization of end-to-end communication overhead
rather than an isolated measurement of firewall processing cost.
The reported RTT measurements include the overhead introduced by virtualization and containerization, which may increase latency and jitter compared to a native deployment~\cite{suo2018analysis}. 

Because \fwname is deployed on the ground segment, its computational requirements are less constrained than those of an onboard deployment.

\subsection{Evaluation Summary}

Our evaluation demonstrates that \fwname\ successfully mitigates
representative attacks across all four classes of our CSP attack
taxonomy, including network identity abuse, management service abuse,
application semantic abuse, and mission context abuse. These results
show that combining CSP-aware packet inspection with mission-specific
policy enforcement enables finer-grained authorization than CSP's
native node and port abstractions alone. \fwname\ achieves this
protection while introducing a median RTT overhead of only 1.42~ms.
Overall, the results demonstrate that \fwname\ can enforce
mission-aware security policies with modest communication overhead.
\section{Related Work}

\textbf{Specialized Firewalls.} FLOWGUARD~\cite{Hu2014Flowguard} introduced a policy-aware firewall for Software-Defined Networks (SDNs) that detects and resolves firewall policy violations caused by dynamic OpenFlow rule updates and changing network state. The framework performs network-wide flow analysis and automatically enforces access control policies as network configurations evolve. In contrast, \fwname\ targets the CubeSat Space Protocol (CSP), enforcing security policies based on CSP network identities, management services, application command semantics, and mission context.

\textbf{CubeSat Space Protocol Security.} Driouch \emph{et al.}~\cite{driouch2024distributed} proposed a distributed intrusion detection system (IDS) for CubeSat networks that performs anomaly detection on CSP-over-CAN communications using deep learning. Their architecture distributes intrusion detection by executing lightweight inference onboard the spacecraft while performing computationally intensive model training at the ground segment. The proposed IDS detects attacks such as denial-of-service, fuzzing, and replay under CubeSat resource constraints. However, the approach operates on CAN traffic characteristics, including packet timing, CAN identifiers, and payload data, making it dependent on the underlying CAN transport rather than the semantics of the CubeSat Space Protocol itself. Consequently, it does not analyze CSP services, application commands, or mission-specific communication policies. In contrast, \fwname\ performs protocol-aware policy enforcement by inspecting CSP packet fields and application semantics, enabling mission-defined access control based on CSP identities, management services, application command semantics, and mission context.

Havermans \emph{et al.}~\cite{havermans2025fuzzing} investigated the security of space communication protocol implementations through grammar-based and coverage-guided fuzzing. Their study evaluated open-source implementations of multiple space protocols, including the official \texttt{libcsp} implementation, identifying implementation vulnerabilities such as memory safety errors and input validation flaws that could be remotely exploited. While this work improves the robustness of protocol implementations by discovering software vulnerabilities, it does not address protocol-level security policies or runtime access control. In contrast, \fwname\ focuses on enforcing mission-defined security policies over CSP communications, preventing unauthorized or unsafe protocol interactions even when the protocol implementation itself is free of software vulnerabilities. 

\section{Discussion}

\subsection{Limitations}
\textbf{CSP Versions.} Our analysis focuses on CSP version 1, which remains widely used in existing CSP deployments and flight software. Although CSP version 2 introduces changes to the packet header and addressing scheme, the fundamental security concerns identified in this work also apply to CSP v2. Therefore, the main conclusions of our security analysis and the policy principles underlying \fwname\ extend to CSP v2.

\textbf{Security Analysis Scope.} Our security analysis primarily examines CSP's identity, trust, service exposure, application semantics, and mission-context assumptions. We do not evaluate the security of cryptographic implementations or conduct cryptographic attacks against CSP. Consequently, vulnerabilities arising from the implementation or configuration of authentication, encryption, or key-management mechanisms are outside the scope of this work.

\subsection{Future Work}

\textbf{CSP Authentication and Authorization.}
A second direction is the development of authentication and authorization mechanisms for CSP. Rather than relying solely on network-level identity and port-based access control, such mechanisms could provide authenticated identities and fine-grained authorization for CSP services and application commands. We plan to investigate an authentication and authorization protocol that can operate on top of CSP while remaining suitable for the resource constraints of CubeSat systems.

\section{Conclusion}

We presented a systematic security analysis of the CubeSat Space Protocol (CSP), identifying security limitations across network identities, management services, application semantics, and mission context. Based on these findings, we developed a taxonomy of CSP attacks and designed \fwname, a modular, mission-aware firewall that enforces policies across these four dimensions. We evaluated \fwname\ using a virtual testbed based on a real-world CubeSat mission, where it successfully blocked all representative attacks while introducing only 1.42~ms of additional median round-trip latency. Our results demonstrate that protocol- and mission-aware policy enforcement provides a practical approach for improving the security of CSP-based spacecraft networks.

\section*{Acknowledgment}
This work was supported in part by the Computing Research Association's
UR2PhD program and by startup funds from New Mexico State University.

\bibliographystyle{IEEEtran}
\bibliography{bib}

\end{document}